\documentclass{PoS}

\usepackage{slashbox}
\usepackage{epsfig}
\usepackage{wrapfig}
\usepackage{lipsum,booktabs}

\newcommand\bef{\begin{figure}}
\newcommand\eef[1]{\label{fg:#1}\end{figure}}
\newcommand\besf{\begin{subfigure}}
\newcommand\eesf[1]{\label{sfg:#1}\end{subfigure}}
\newcommand\beq{\begin{equation}}
\newcommand\eeq[1]{\label{#1}\end{equation}}
\newcommand\beqa{\begin{eqnarray}}
\newcommand\eeqa[1]{\label{#1}\end{eqnarray}}
\newcommand\bet{\begin{table}}
\newcommand\eet[1]{\label{tb:#1}\end{table}}
\newcommand\best{\begin{subtable}}
\newcommand\eest[1]{\label{stb:#1}\end{subtable}}
\newcommand\betb{\begin{center}\begin{tabular}}
\newcommand\eetb{\end{tabular}\end{center}}
\newcommand\beit{\begin{itemize}}
\newcommand\eeit{\end{itemize}}

\newcommand\fgn[1]{Figure \ref{fg:#1}}

\newcommand\tbn[1]{Table \ref{tb:#1}}

\newcommand\incfig[2]{\includegraphics[scale=#1]{#2}}

\title{Spectroscopy of charmed baryons from lattice QCD}

\ShortTitle{Spectroscopy of charmed baryons from lattice QCD}

\author{\speaker{M. Padmanath} \\%
        Institute of Physics, University of Graz, 8010 Graz, Austria.\\
        E-mail: \email{padmanath.madanagopalan@uni-graz.at}}

\author{Robert\ G.\ Edwards\\
        Jefferson Laboratory, 12000 Jefferson Avenue, Newport News, VA 23606, USA\\
        E-mail: \email{edwards@jlab.org}}

\author{Nilmani\ Mathur\\
        Department of Theoretical Physics, Tata Institute of Fundamental Research, Mumbai, India\\
        E-mail: \email{nilmani@theory.tifr.res.in}}

\author{Michael\ Peardon\\
        School of Mathematics, Trinity College Dublin, Ireland\\
        E-mail: \email{mjp@maths.tcd.ie}}

\abstract{We present the ground and excited state spectra of singly, doubly and triply charmed 
baryons by using dynamical lattice QCD.  A large set of baryonic operators that respect the 
symmetries of the lattice and are obtained after subduction from their continuum analogues are 
utilized. Using novel computational techniques correlation functions of these operators are 
generated and the variational method is exploited to extract excited states. The lattice spectra 
that we obtain have baryonic states with well-defined total spins up to $\frac{7}{2}$ and the 
low lying states remarkably resemble the expectations of quantum numbers from SU(6)$\times$ 
O(3) symmetry. Various energy splittings between the extracted states, including splittings due 
to hyperfine as well as spin-orbit coupling, are considered and those are also compared against 
similar energy splittings at other quark masses.}

\FullConference{The 32nd International Symposium on Lattice Field Theory,\\
		23-28 June, 2014\\
		Columbia University New York, NY}

\begin{document}

\vspace{-0.5cm}
\section{Introduction}
\vspace{-0.3cm}
There is a lot of resurgent scientific interest in heavy hadron spectroscopy with the exciting 
experimental observations during the past decade at CLEO-c, BaBar, the Tevatron, Belle, BES and LHCb.
This interest continues, anticipating the observations from large statistical samples that will be 
collected in many ongoing and future experiments, like  BESIII, Belle II at KEK, experiments at 
the LHCb, and the planned PANDA experiment at GSI/FAIR. However, in contrast to the heavy quarkonia 
spectroscopy, which have been getting extensive scientific attention in both experiments and 
theoretical studies, the heavy baryons have not been explored in great detail, though they also can 
aid to augment our understanding about the strong interaction. 
So far, only a few singly charmed
baryons have been discovered \cite{PDG}, the experimental status of the doubly charm baryon is
controversial \cite{PDG} and there are no observations yet for triply charm baryons.
Moreover, quantum number assignment has not been made for most of these observed states. In light 
of these existing and future experimental prospects, it is highly desirable to have model 
independent calculations of heavy baryon spectra from first principles calculations, such as from 
lattice QCD. Such calculations will naturally provide crucial inputs to the future experimental 
discovery and can also provide a guide in identifying the unknown quantum numbers of the 
experimentally discovered states, based on what one expects from QCD. Furthermore, such lattice results 
can also be compared with those obtained from potential models \cite{Crede:2013kia}, which have 
been very successful in the case of heavy mesons.  
Until very recently, lattice QCD results,  including quenched \cite{Lewis:2001iz, Mathur:2002ce} as 
well as full QCD \cite{Basak:2012py,Durr:2012dw,Namekawa:2013vu,Alexandrou:2012xk,Briceno:2012wt,Bali:2012ua}, 
on charmed baryons included only the ground states with spin up to ${3\over 2}$. 
In this proceeding, we present our results 
on comprehensive excited state spectra of singly, doubly and triply charmed baryons with spin up to 
7/2 for both parities \cite{Padmanath:2013bla}. 

\vspace{-0.5cm}
\section{Numerical details}
\vspace{-0.3cm}
These calculations employed the ensemble of dynamical anisotropic gauge field configurations generated 
by the Hadron Spectrum Collaboration (HSC) to extract the highly excited hadron spectra.
With a large anisotropy co-efficient, $\xi=a_s/a_t=3.5$, we could achieve $a_t~m_c\ll1$ and hence use 
the standard relativistic formulation of Fermions for all the quark flavors from light to charm.
The gauge configurations used, were generated with the Symanzik-improved gauge action and $N_f=$2+1
Fermionic fields in the sea, described using an anisotropic clover action with tree-level tadpole improvement
and stout-smeared spatial links. The temporal lattice spacing, $a_t^{-1}=5.67$GeV, was 
determined by equating the lattice estimate of $m_{\Omega}$ to its physical value. With a spatial 
lattice extension of $\sim$1.9 ~fm, we expect the finite size effects on our spectra to be much 
less than that for the light hadron spectra. More details of the formulation of actions as well as the 
techniques used to determine the anisotropy parameters can be found in Refs. \cite{Edwards:2008ja, Lin:2008pr}. 
\bet[h]
\centering
\betb{cccc|cc|cc}
\hline
Lattice size & $a_t m_\ell$   & $a_t m_s$ & $N_{\mathrm{cfgs}}$ &  $m_\pi/$MeV & $a_t m_\Omega$ & $N_{\mathrm{tsrcs}}$ & $N_{\mathrm{vecs}}$\\\hline
$16^3 \times 128$ & $-0.0840$ & $-0.0743$ & 96 & 391        & $0.2951(22)$   & 4 & 64   
\\\hline
\eetb
\caption{Details of the gauge-field ensemble used. $N_{\mathrm{cfgs}}$  is the number of gauge-field 
configurations, while $N_{\mathrm{tsrcs}}$ and $N_{\mathrm{vecs}}$ are the number of time sources per 
configuration and the number of distillation eigenvectors used for each time source, respectively.}
\eet{lattices}
Lattice specifications of these gauge field ensembles used in this work are given in \tbn{lattices}.

\vspace{-0.4cm}
\section{Operator construction and analysis}
\vspace{-0.2cm}
\begin{wraptable}{r}{5.5cm}
\betb{c | c | c | c } 
\hline
                              &   $G_1$  &     H    &    $G_2$     \\ \hline
$\Omega_{ccc}$                &    20    &     33   &     12       \\\hline
$\Omega_{cc}$, $\Xi_{cc}$     &          &          &              \\
$\Omega_{c}$, $\Sigma_c$      &    55    &     90   &     35       \\\hline
$\Lambda_c$                   &    53    &     86   &     33       \\\hline
$\Xi_{c}$                     &   116    &    180   &     68       \\\hline
\eetb\vspace{-0.2cm}
\caption{Total number of operators constructed for various charm baryons in each lattice irrep.}
\label{totoperators}
\end{wraptable}
Hadron spectroscopy on the lattice proceeds through the computation of the Euclidean two point 
correlation functions, 
\beq
C_{ij}(t_f-t_i) = \langle 0|O_j(t_f)\bar{O}_i(t_i)|0\rangle = \sum_{n} {Z^{n*}_iZ^n_j\over 2 m_n} e^{-m_n(t_f-t_i)}  
\eeq{2pt} 
for different creation ($\bar{O}_i(t_i)$) and annihilation ($O_j(t_f)$) operators that carry the quantum 
numbers of our interest. The $Z^{n}_i = \langle 0|O_i^{\dagger}|n\rangle$, called the overlap factor, 
carries the information about the quantum numbers of the physical state, $n$. Employing derivative-based 
operator construction formalism \cite{Edwards:2011jj} and using up to two derivatives, we construct a 
large basis of charm baryon operators
that transforms according to the symmetries of the lattice. In Table \ref{totoperators},
we tabulate the total number of operators that forms the basis for the three different lattice irreps. We 
identify a subset of operators (\tbn{operators}) that are formed by considering only the upper two 
components of the Dirac spinor as non-relativistic operators as they form the whole set of operators 
(with $SU(6)\otimes O(3)$) in the non-relativistic limit. We also identify another subset of two 
derivative operators (\tbn{operators}), which are proportional to the field strength tensor, as 
`hybrid' operators \cite{Dudek:2012ag}. For each lattice irrep, we construct $N\times N$ matrix of 
correlation functions, where $N$ is the number of the operators used in the respective irrep. For 
these computations we used advanced smearing technique called `Distillation', so as to reduce the 
computational requirements to a practical level \cite{Peardon:2009gh}.  Further, we employ a variational 
method \cite{Dudek:2007wv} for the extraction of the physical states from these matrix of Euclidean 
correlation functions using the large basis of interpolators and utilize the overlap factors, $Z^{n}_i$, to identify 
the quantum numbers of the extracted physical states \cite{Dudek:2007wv}. Thus we are able to reliably 
extract charm baryon states with spin up to $J=7/2$ for both parities.
\begin{table}[h]
\small
\betb{c | c  | c | c | c } 
\hline
\hline
& $\Omega_{ccc}$ & $\Omega_{cc}$, $\Xi_{cc}$, $\Omega_{c}$, $\Sigma_c$ & $\Lambda_c$ & $\Xi_c$  \\ \hline
D                    & $1/2$~ $3/2$~ $5/2$ ~  $7/2$ & $1/2$ ~  $3/2$ ~  $5/2$ ~  $7/2$ & $1/2$ ~  $3/2$ ~  $5/2$ ~  $7/2$ & $1/2$ ~  $3/2$ ~  $5/2$ ~  $7/2$ \\ \hline
0                    &   0~~~~ 1   ~~~~0   ~~~~ 0   &   1   ~~~~ 1   ~~~~ 0   ~~~~ 0   &   1   ~~~~ 0   ~~~~ 0   ~~~~ 0   &  ~~ 2 ~~~~~ 1 ~~~~~~0 ~~~~~~~0~   \\ \hline
1                    &   1~~~~ 1   ~~~~0   ~~~~ 0   &   3   ~~~~ 3   ~~~~ 1   ~~~~ 0   &   3   ~~~~ 3   ~~~~ 1   ~~~~ 0   &  ~~ 6 ~~~~~ 6 ~~~~~~2 ~~~~~~~0~   \\ \hline
2$_h$                &   1~~~~ 1   ~~~~0   ~~~~ 0   &   3   ~~~~ 3   ~~~~ 1   ~~~~ 0   &   3   ~~~~ 3   ~~~~ 1   ~~~~ 0   &  ~~ 6 ~~~~~ 6 ~~~~~~2 ~~~~~~~0~   \\ \hline
2                    &   2~~~~ 3   ~~~~2   ~~~~ 1   &   6   ~~~~ 8   ~~~~ 5   ~~~~ 2   &   6   ~~~~ 7   ~~~~ 5   ~~~~ 1   &   12  ~~~~ 15  ~~~ 10 ~ ~~~~ 3   \\ \hline
\eetb
\label{nroperators}
\caption{The description of the non-relativistic operators used. D stands for the number of derivatives.
The fractions in the second row are the continuum spin from which the operators are derived from. 
Subscript $h$ in the fifth row stands for the hybrid operators.}
\vspace{-0.2cm}
\eet{operators}


\vspace{-0.5cm}
\section{Results}\vspace{-0.3cm}
We will present our results starting with triply charmed, followed by doubly charmed and then for singly charmed baryons.
In \fgn{ccc_spectrum}(a), we show the spin identified spectra of the triply charmed baryons with reference to 
$3/2$ times the mass of $\eta_c$ to account for the difference in the charm quark content \cite{Padmanath:2013bla}.
The energy splittings are preferable than the absolute value of the energy of the physical states, as it reduces the 
systematic uncertainties due to the quark mass tuning errors, the discretization effects, and also lessen the 
effect of ambiguity in the scale setting procedure. We emphasize the states in the excited bands with relatively 
large overlap onto the nonrelativistic operators by enclosing them in magenta closed curves. These states should 
thus be well described within a nonrelativistic quark model. One can immediately see the extracted spectrum resemble 
expectations based on a model with nonrelativistic quark spins and hence provide a clear signature of 
SU(6)$\otimes$O(3) symmetry in the spectra. A few boxes with thick border correspond to those with a greater 
overlap onto the `hybrid' operators, which might consequently be hybrid states \cite{Dudek:2012ag}.

\bef[t]
\small
\vspace{-0.5cm}
\parbox{.45\linewidth}{
\centering
  \includegraphics[width=70mm,trim=0 0 0 0 mm, clip=true]{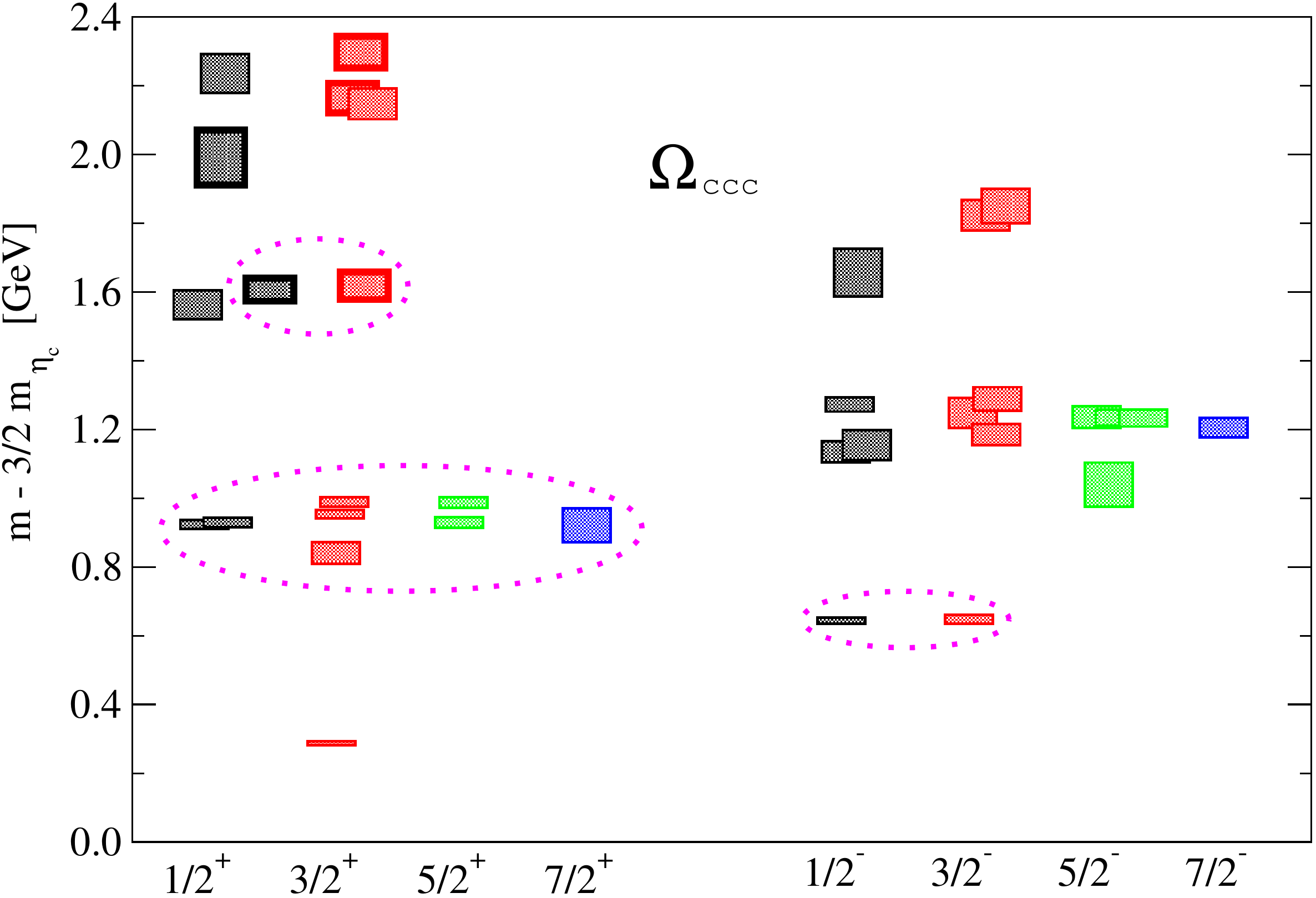}\\
(a)}
\hspace{0.75cm}
\parbox{.45\linewidth}{ 
\centering
  \includegraphics[width=70mm,trim=0 0 0 1 mm, clip=true]{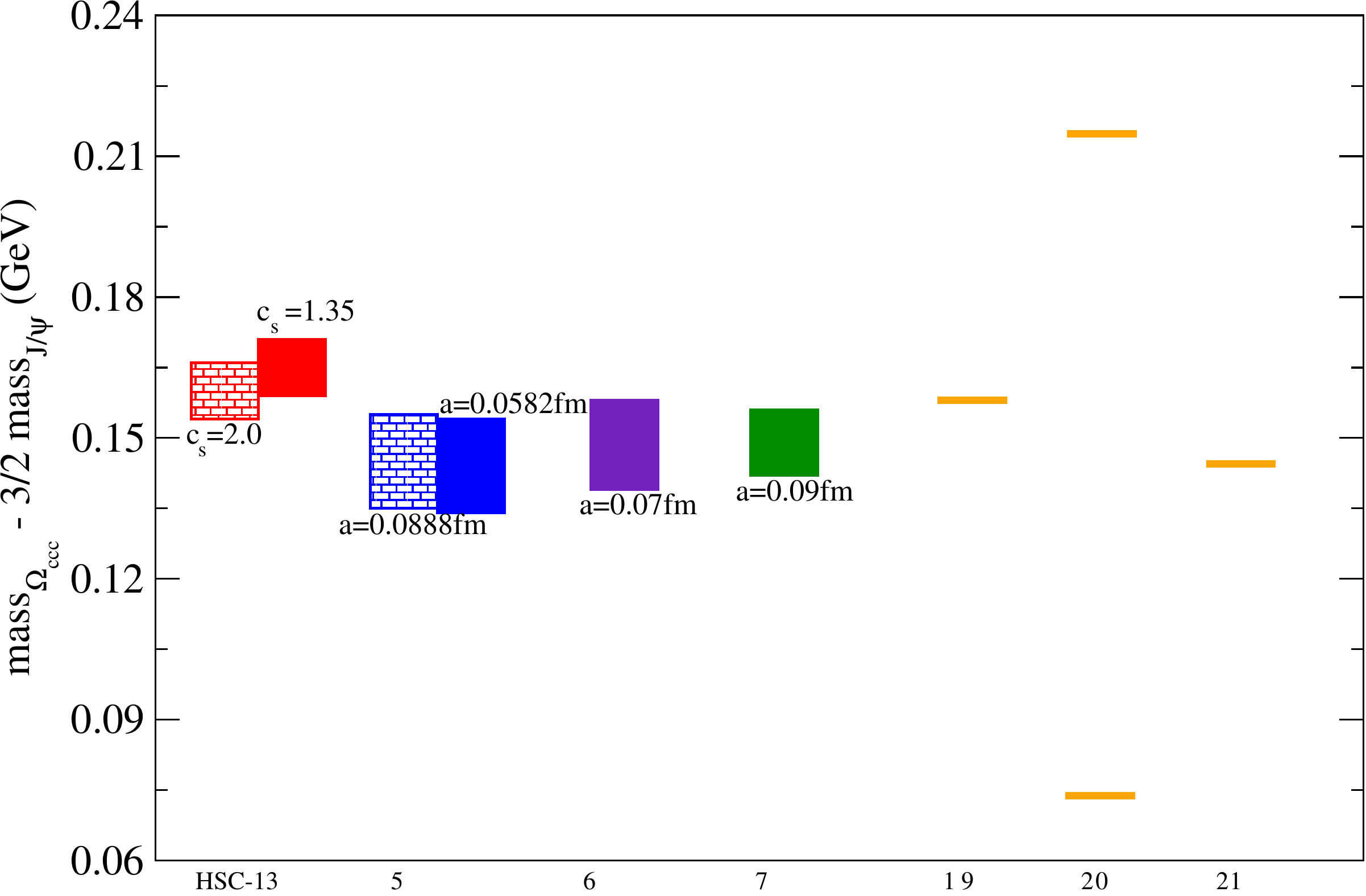}\\
(b)}\vspace{-0.2cm}
\caption{(a) Spin identified spectra of triply-charmed baryons with respect to $\frac32 m_{\eta_c}$ mass. 
The states in the excited bands inside the magenta ellipses are those with relatively large overlap to 
non-relativistic operators. The boxes with thick borders corresponds to the states with strong overlap with 
hybrid operators. (b) Mass splitting of the ground state of $J^P=\frac32^+ ~\Omega_{ccc}$ from $\frac32$ 
times the mass of $J/\psi$ meson is compared for various lattice calculations \cite{Basak:2012py,Durr:2012dw,Namekawa:2013vu}
and potential model calculations \cite{Martynenko:2007je,Migura:2006ep,Hasenfratz:1980ka}. }
\eef{ccc_spectrum}

A second calculation was carried out with a boosted value for the improvement co-efficient ($c_s=2.0$) in the 
charm quark action from its tree level value ($c_s=1.35$), so as to assess the effect of the radiative corrections,
which could lead to significant change in the physical predictions. As was observed in the study of charmonium 
spectrum \cite{Liu:2012ze}, we observe a positive shift of approximately 45MeV in the spectrum with reference 
to $3/2m_{\eta_c}$\cite{Padmanath:2013bla}. In \fgn{ccc_spectrum}(b), we plot our estimates for 
$m_{\Omega_{ccc}}-3/2m_{J/\psi}$ from these two calculations with the boosted value for the improvement co-efficient 
and its tree level value respectively. In the same plot, we also compare these estimates with other lattice 
calculations~\cite{Basak:2012py,Durr:2012dw,Namekawa:2013vu}, which use different discretization and so have 
distinct artefacts. We also compare these estimates with some potential model calculations~\cite{Martynenko:2007je,
Migura:2006ep,Hasenfratz:1980ka}. Consistency of our estimates with other lattice calculations gives confidence in our results.

Spin dependent energy splittings provide important insights into the nature of the interactions within the 
physical states. The most notable baryon energy splittings are those due to spin-orbit coupling and 
the hyperfine splittings. In Figure \ref{ccc_socoup}, we show the the absolute values of energy splittings between 
the physical states, which originate from the spin-orbit interaction of the following (L, S) pairs : 
(2,3/2-in the left), (2,1/2-in the middle) and (1,1/2-in the right column). The plot contains these splittings 
for triple flavored baryons at varying quark masses from {\it light} to {\it bottom}. We identified these (L,S) 
pairs by finding the operators, which incorporate these pairs, that have major overlaps to these states. While 
the data at the charm quark mass is from 
this work, the data for {\it bottom} baryons are obtained from Ref. \cite{Meinel:2012qz} and data at the 
{\it light} and {\it strange} quark masses are from 
\begin{wrapfigure}{r}{80mm}
\begin{center}
  \incfig{0.32}{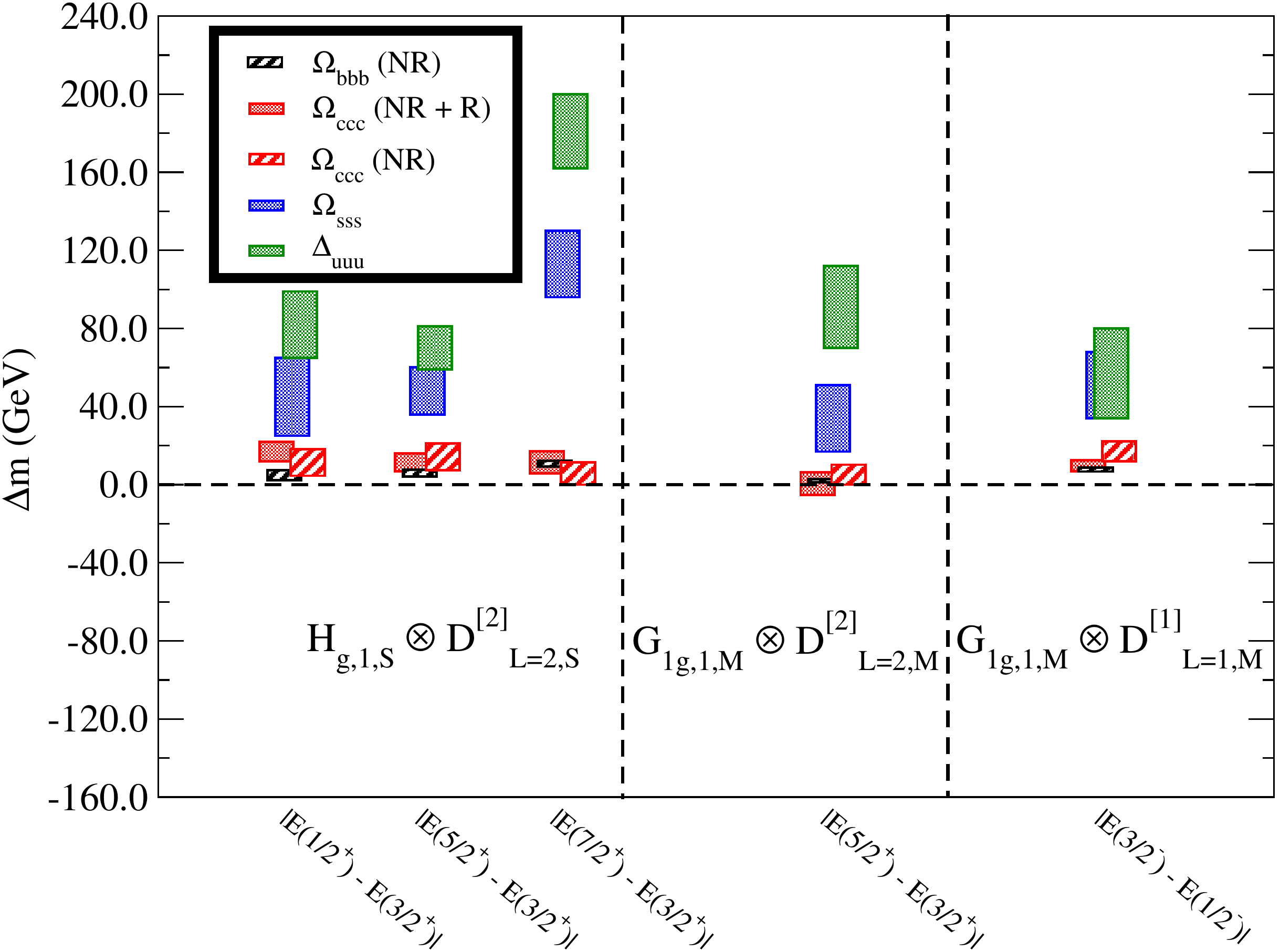}\\
\caption{ Energy splittings between states with same L and S values, starting from light to heavy 
triple-flavoured baryons. For $\Omega_{bbb}$, results are with only non-relativistic operators 
\cite{Meinel:2012qz}. For $\Omega_{ccc}$, results from relativistic and non-relativistic as well as 
only non-relativistic operators are shown, and for the light and strange baryons results are with 
relativistic and non-relativistic operators \cite{Edwards:2011jj}.}
\end{center}
\label{ccc_socoup}
\vspace*{-0.1in}
\end{wrapfigure}
Ref. \cite{Edwards:2011jj}. It is very clear from the plot 
that these splittings are very near to zero at both {\it charm} and {\it bottom} quark masses and thus indicates 
the non-relativistic nature of heavy baryons. 
However, data with higher statistics is necessary for precise determination of the heaviness of charm quark.

Next, we show the spin identified spectra of doubly charmed baryons \cite{Padmanath:2013bla} in \fgn{ccq_spectrum}. 
The spectra are plotted with the mass of $\eta_c$ subtracted from them so as to account for valence charm quark content. 
The states with strong overlap with the nonrelativistic operators are emphasized by enclosing them within magenta
closed curves, while the states with hybrid nature are shown with thick border. As is evident from the plots, 
the spectra shows excellent agreement between the number of states in the lower nonrelativistic bands and 
the expectations as per a model with SU(6)$\otimes$O(3) symmetry. Further, incorporating the doubly flavored baryon data, 
starting from {\it light} to {\it bottom}, we studied the quark mass dependence of various energy splittings and 
using heavy quark effective theory motivated model, we were able to predict the following at the bottom sector : 
$m_{B_c^*}-m_{B_c} = 80\pm8$ MeV and $m_{\Omega_{ccb}} = 8050\pm10$ MeV.

\bef[th]
\small
\parbox{.5\linewidth}{
\centering
  \includegraphics[scale=0.3]{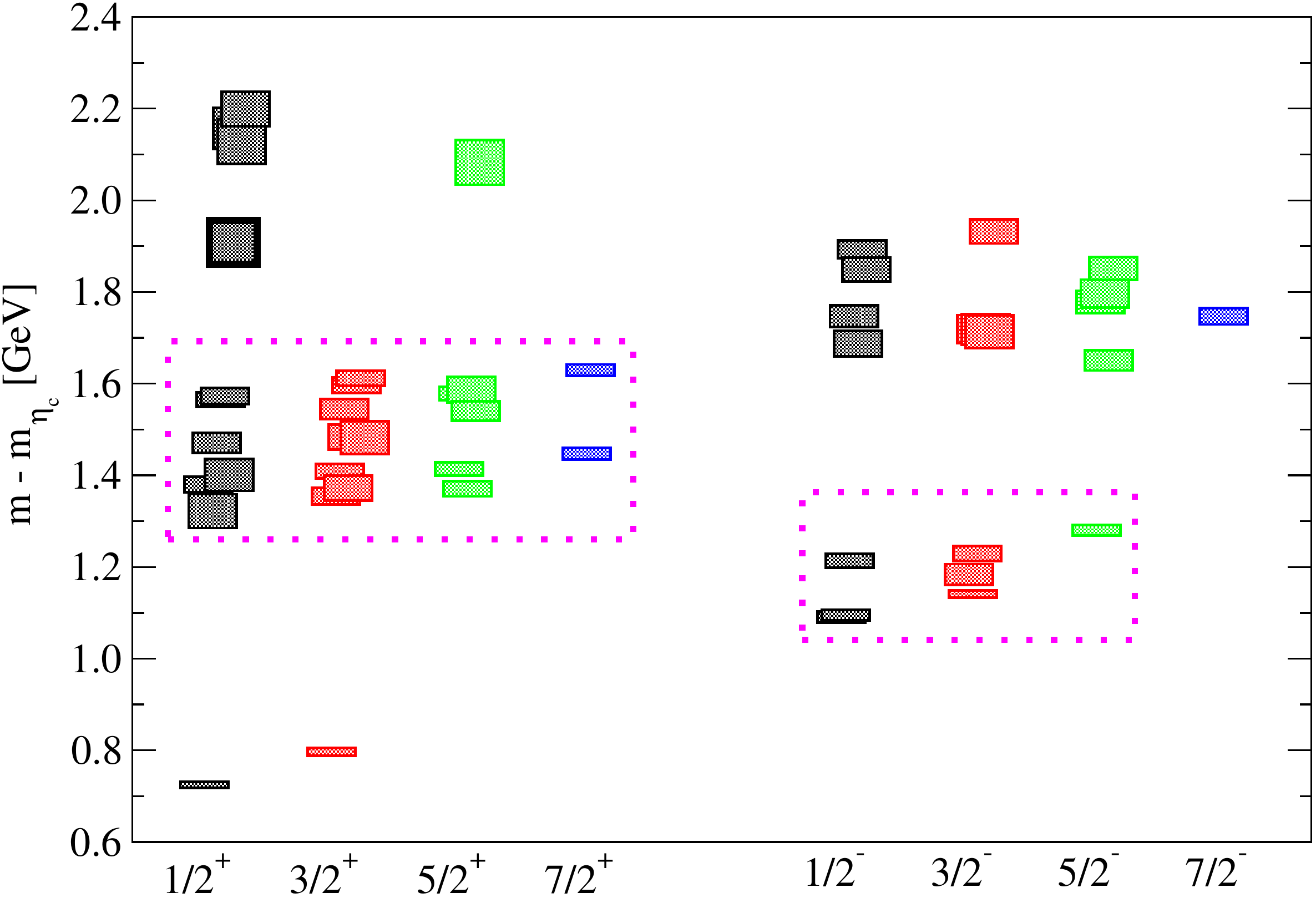}\\
(a)}
\parbox{.5\linewidth}{ 
\centering
  \includegraphics[scale=0.3]{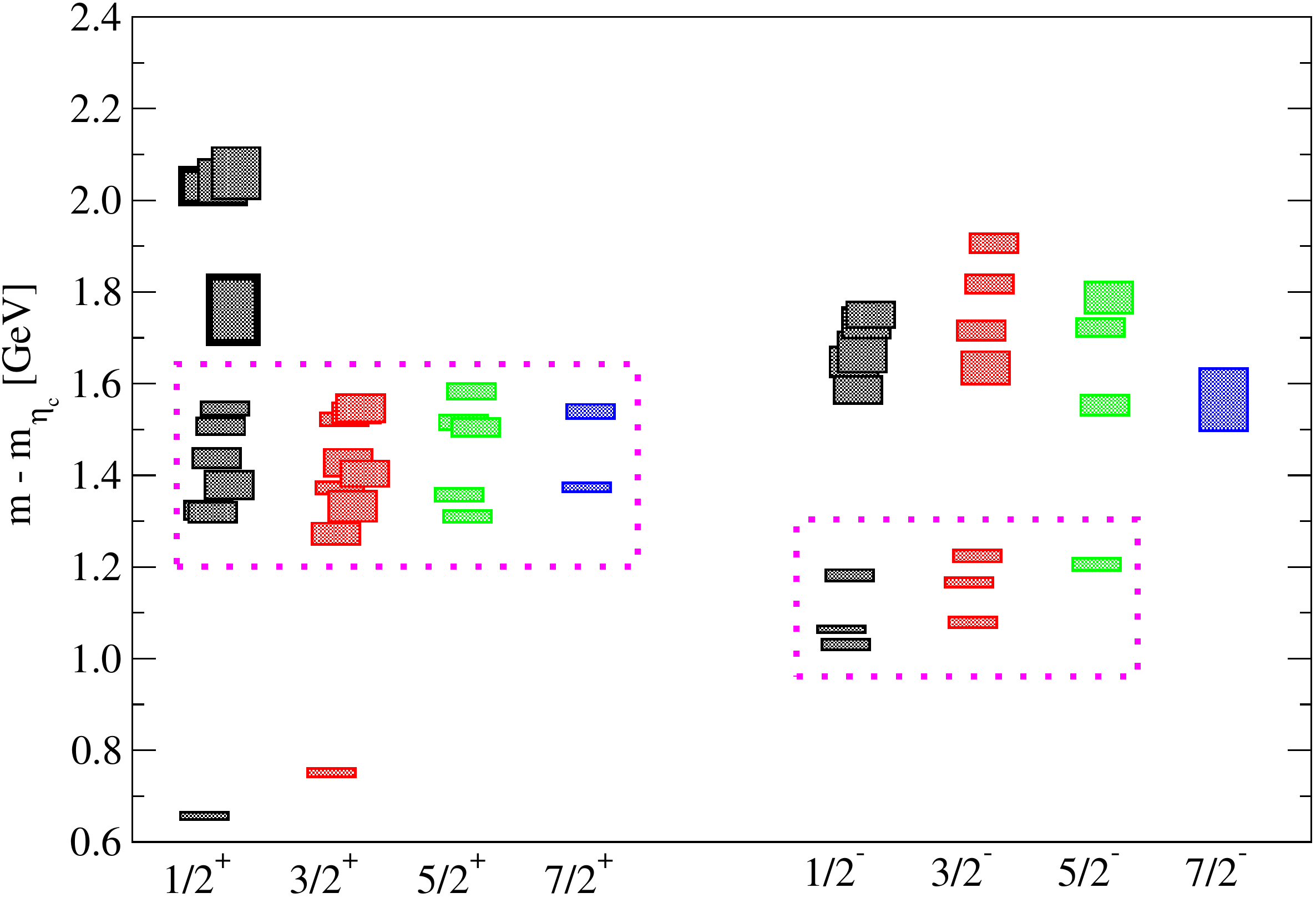}\\
(b)}
\caption{ Spin identified spectra of (a) $\Omega_{cc}$ and (b) $\Xi_{cc}$ baryon for both parities 
and with spins up to $\frac72$ {\it w.r.t.} $m_{\eta_c}$. The keys are same as in Figure 1(a).}
\eef{ccq_spectrum}

\bef[th]
\small
\vspace{-0.5cm}
\parbox{.5\linewidth}{
\centering
  \includegraphics[scale=0.3]{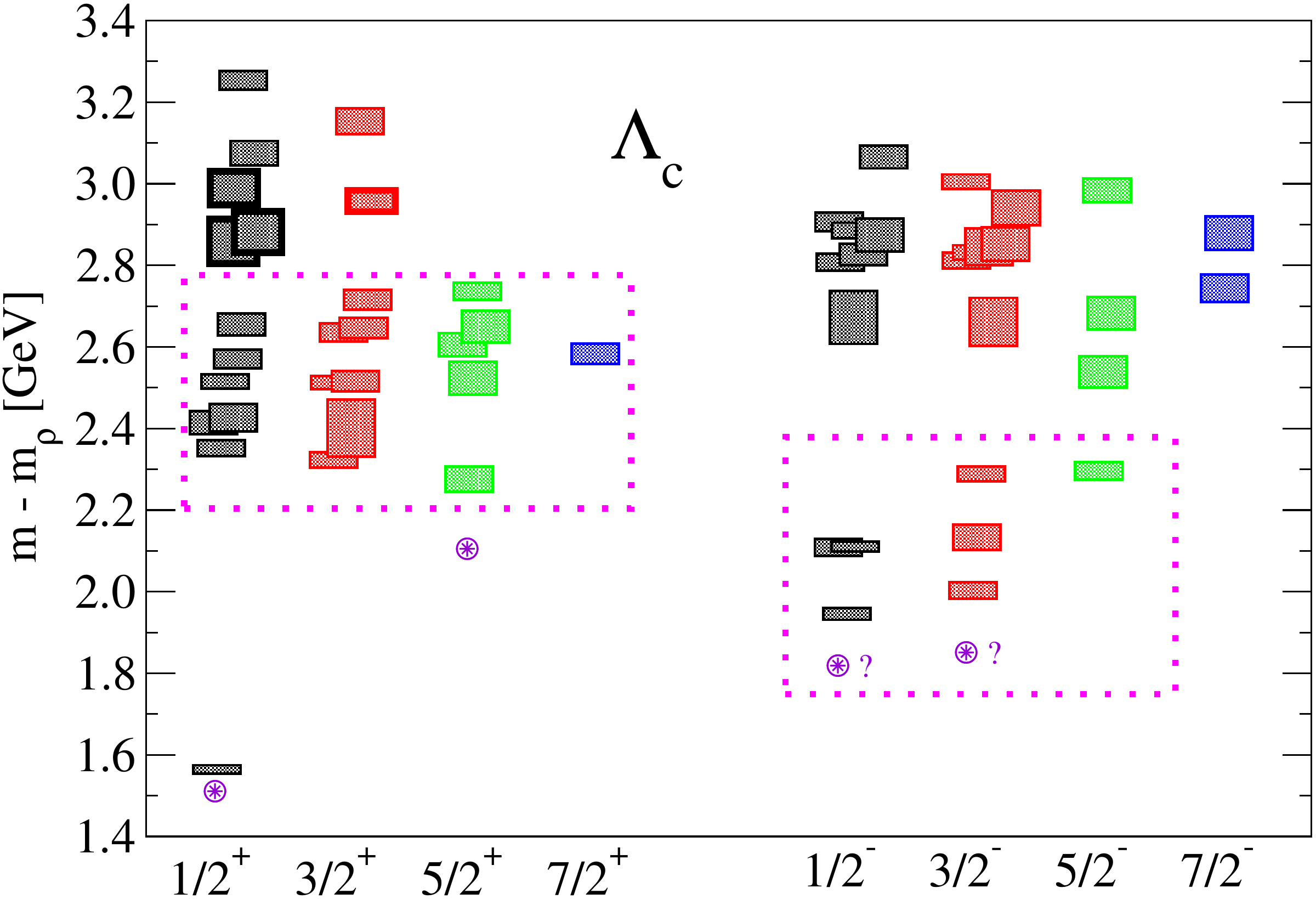}\\
(a)}
\parbox{.5\linewidth}{ 
\centering
  \includegraphics[scale=0.3]{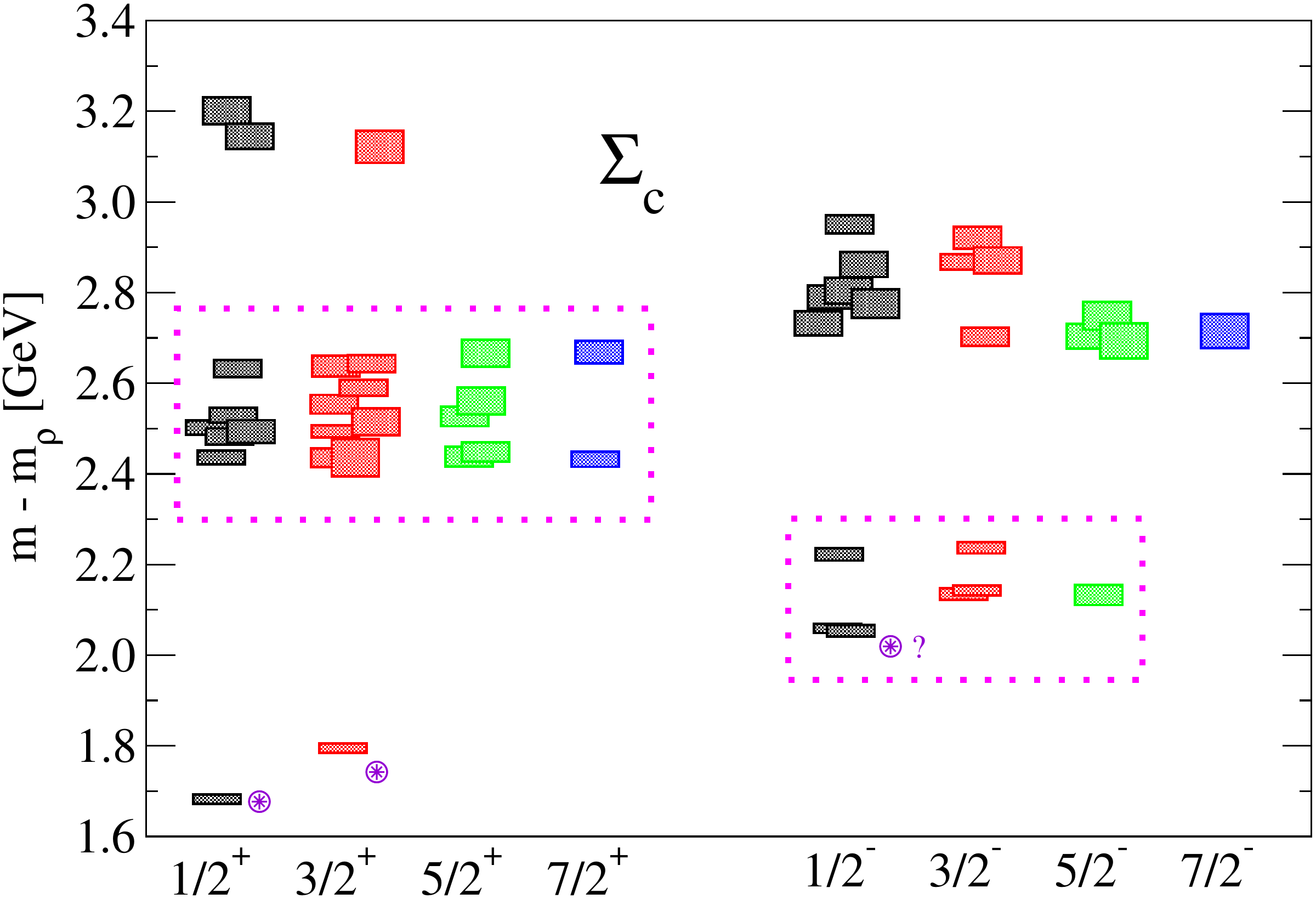}\\
(b)}
\parbox{.5\linewidth}{
\centering
  \includegraphics[scale=0.3]{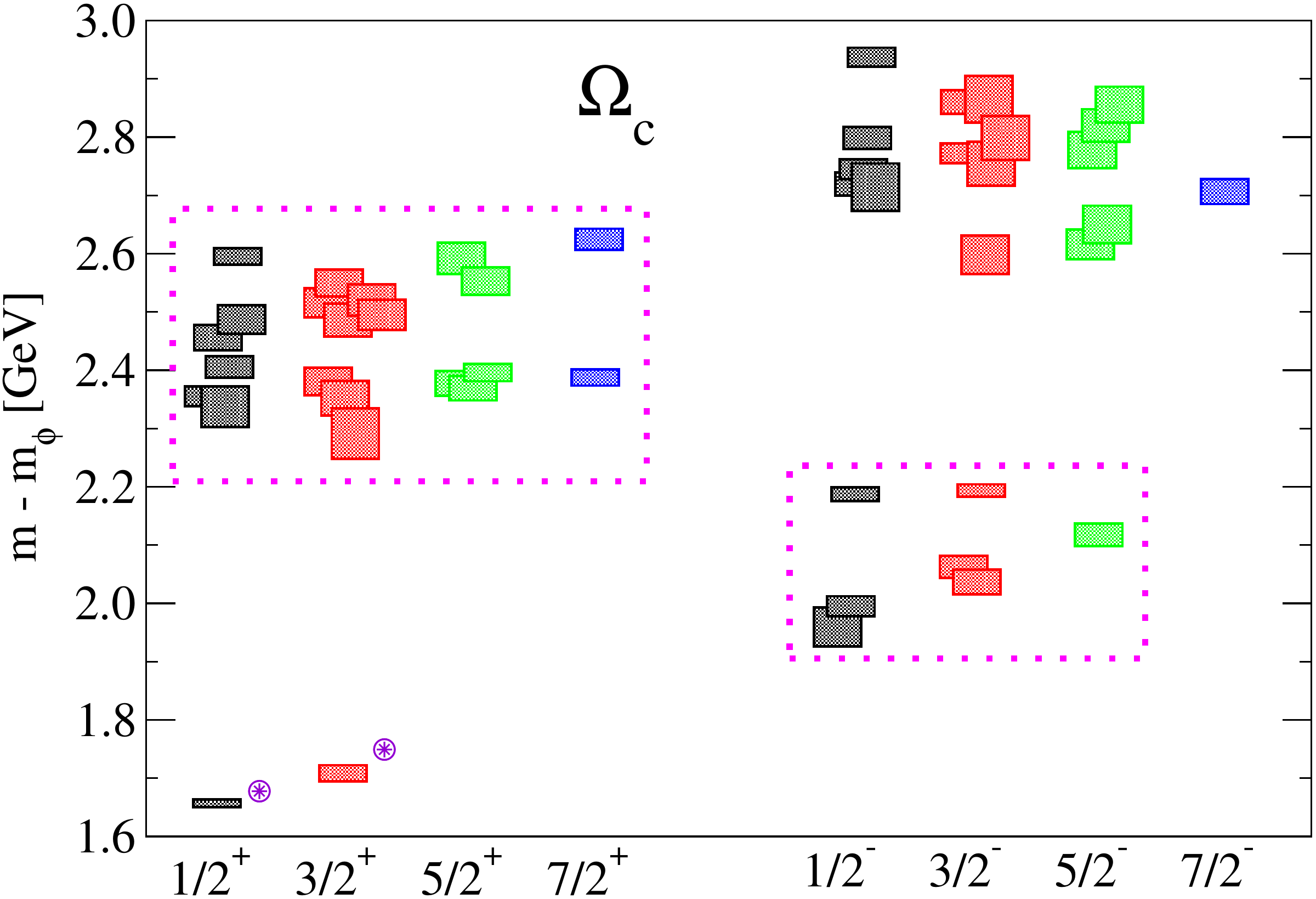}\\
(c)}
\parbox{.5\linewidth}{ 
\centering
  \includegraphics[scale=0.3]{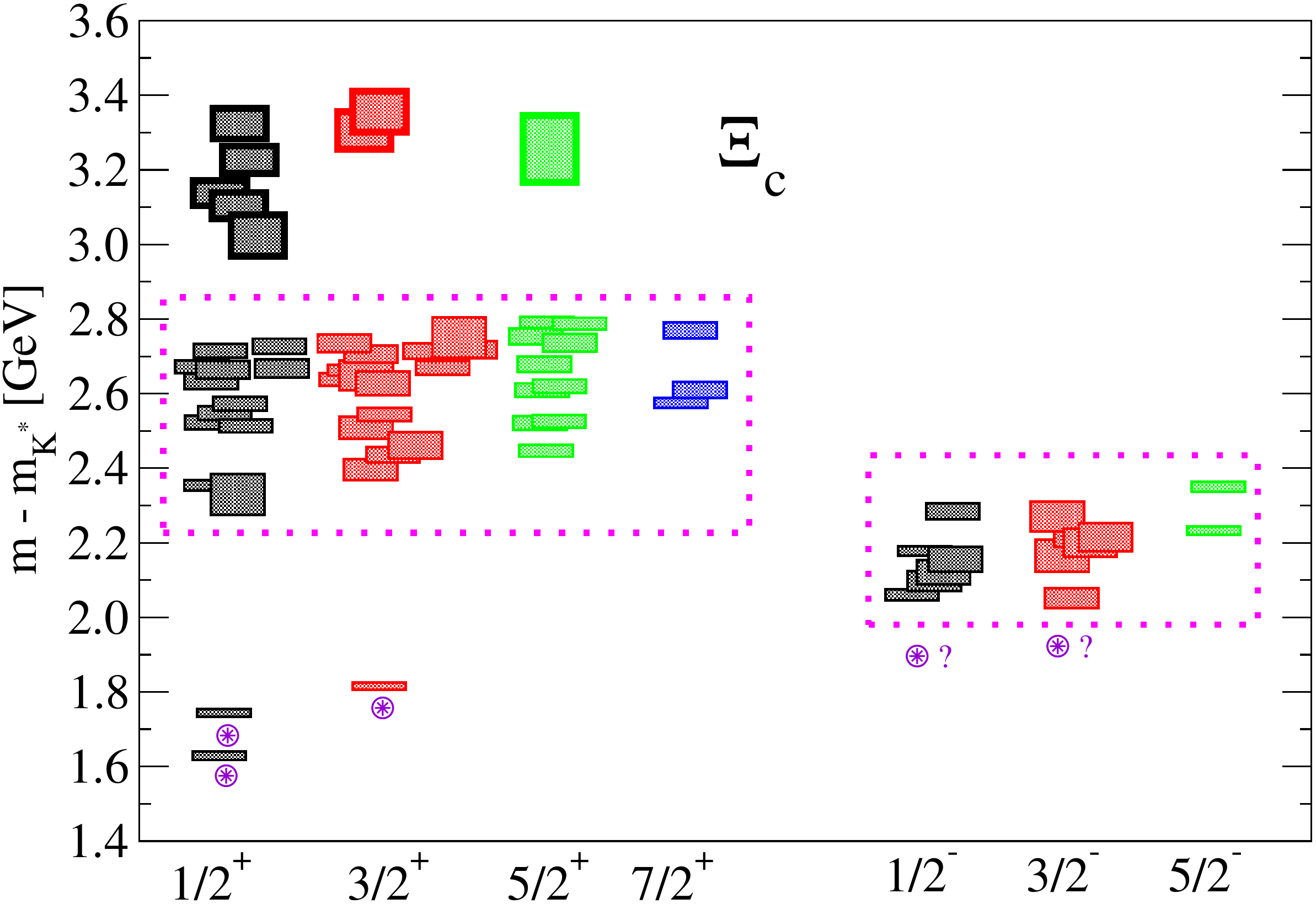}\\
(d)}\vspace{-0.2cm}
\caption{The spin identified spectra of (a) $\Lambda_{c}$, (b) $\Sigma_{c}$, 
(c) $\Omega_{c}$ and (d) $\Xi_{c}$ baryons for both parities in terms of the splitting from 
the respective vector meson. The keys are same as in Figure 1(a).}
\eef{LS_spectrum}

In \fgn{LS_spectrum}, we show the spin identified spectra of the singly charmed baryons, which include
$\Lambda_c$, $\Sigma_c$, $\Xi_c$ and $\Omega_c$. The spectra are shown in terms of the splitting from 
the respective vector meson, such that all the singly charm baryon spectra shown carry an 
effective single valence charm quark content and hence all the spectra have the same leading systematic corrections. 
The circled stars in violet shows the experimental candidates. Hence the leading corrections in the 
observed discrepancies between the experimental data and our estimates could be attributed to the 
charm quark tuning, the discretization of the charm quark action and the unphysically heavy light pion mass. 
Nevertheless, one can see a good agreement for the singly charmed baryon spectra also, between the number of 
states in the lower non-relativistic bands and the expectations as per a model with SU(6)$\otimes$O(3) symmetry.

\vspace{-0.4cm}
\section{Conclusions}
\vspace{-0.2cm}
In this work we present the first calculation of the ground and excited state spectra of singly, doubly 
and triply-charmed baryons using dynamical lattice QCD. Employing state-of-the-art techniques like 
derivative-based operator construction formalism, `distillation' and variational fitting method, we 
are able to reliably extract spectra of charm baryons with well-defined total spin up to 7/2 for both parities. The low lying 
states in all the spectra remarkably resemble the expectations based on a model with SU(6) $\otimes$ O(3) 
symmetry. We study quark mass dependence of various energy splittings including those that originate 
from hyperfine interactions, as well as spin-orbit interactions. From these studies, we also make the predictions 
in the \textit{bottom} sector which are $m_{B_c^*}-m_{B_c} = 80\pm8$ MeV and $m_{\Omega_{ccb}} = 8050\pm10$ MeV.
However, it is to be noted that the systematic uncertainties like chiral extrapolation, discretization effects 
and finite volume effects are not quantitatively addressed here. 
Further, one needs to incorporate multi-hadron operators in the full
analysis so as to assess the extent of their influence in the above conclusions and for precise quantitative 
description of these resonance states.

\vspace{-0.4cm}
\section{Acknowledgements}
\vspace{-0.1cm}
We thank our colleagues within the Hadron Spectrum Collaboration.  Chroma~\cite{Edwards:2004sx} and 
QUDA\\\cite{Clark:2009wm,Babich:2010mu} were used to perform this work on the Gaggle and Brood clusters 
of the DTP, TIFR, Mumbai and at Lonsdale cluster maintained by the Trinity Centre for High Performance 
Computing and at Jefferson Laboratory. MP acknowledges support from the Austrian Science Fund (FWF):[I1313-N27], 
the Trinity College Dublin Indian Research Collaboration Initiative and the CSIR, India for financial 
support through the SPMF. RGE acknowledges support from the U.S. Department of Energy and Jefferson 
Science Associates which manages and operates Jefferson Laboratory.

\end{document}